\documentclass[journal]{IEEEtran}
\usepackage{cite}
\ifCLASSINFOpdf
   \usepackage[pdftex]{graphicx}
\else
   \usepackage[dvips]{graphicx}
\fi
\usepackage[cmex10]{amsmath}
\usepackage{amssymb}

\DeclareMathOperator*{\argmin}{\arg\!\min}
\usepackage{color}
\begin{document}
%
% paper title
% can use linebreaks \\ within to get better formatting as desired
% Do not put math or special symbols in the title.
\title{Out-of-Band Power Reduction in NC-OFDM with Optimized Cancellation Carriers Selection}
%
%
% author names and IEEE memberships
% note positions of commas and nonbreaking spaces ( ~ ) LaTeX will not break
% a structure at a ~ so this keeps an author's name from being broken across
% two lines.
% use \thanks{} to gain access to the first footnote area
% a separate \thanks must be used for each paragraph as LaTeX2e's \thanks
% was not built to handle multiple paragraphs
%

\author{Pawel~Kryszkiewicz,~\IEEEmembership{Student Member,~IEEE}
        and~Hanna~Bogucka,~\IEEEmembership{Senior Member,~IEEE,}

\thanks{The work presented in this paper has been supported by the European 7th Framework Programme project NEWCOM\#, contract no. 318306.}
\thanks{
The authors are with the Chair of Wireless Communications at Poznan University of Technology, Poland. }
\thanks{Copyright (c) 2013 IEEE. Personal use is permitted. For any other purposes, permission must be obtained from the IEEE by emailing pubs-permissions@ieee.org.
This is the author's version of an article that has been published in this journal. Changes were made to this version by the publisher prior to publication.
The final version of record is available at http://dx.doi.org/10.1109/LCOMM.2013.081813.131515}
}

% note the % following the last \IEEEmembership and also \thanks - 
% these prevent an unwanted space from occurring between the last author name
% and the end of the author line. i.e., if you had this:
% 
% \author{....lastname \thanks{...} \thanks{...} }
%                     ^------------^------------^----Do not want these spaces!
%
% a space would be appended to the last name and could cause every name on that
% line to be shifted left slightly. This is one of those "LaTeX things". For
% instance, "\textbf{A} \textbf{B}" will typeset as "A B" not "AB". To get
% "AB" then you have to do: "\textbf{A}\textbf{B}"
% \thanks is no different in this regard, so shield the last } of each \thanks
% that ends a line with a % and do not let a space in before the next \thanks.
% Spaces after \IEEEmembership other than the last one are OK (and needed) as
% you are supposed to have spaces between the names. For what it is worth,
% this is a minor point as most people would not even notice if the said evil
% space somehow managed to creep in.

% The paper headers
\markboth{IEEE Communications Letters}%
{Shell \MakeLowercase{\textit{et al.}}: Bare Demo of IEEEtran.cls for Journals}
% The only time the second header will appear is for the odd numbered pages
% after the title page when using the twoside option.
% 
% *** Note that you probably will NOT want to include the author's ***
% *** name in the headers of peer review papers.                   ***
% You can use \ifCLASSOPTIONpeerreview for conditional compilation here if
% you desire.

% If you want to put a publisher's ID mark on the page you can do it like
% this:
%\IEEEpubid{0000--0000/00\$00.00~\copyright~2012 IEEE}
% Remember, if you use this you must call \IEEEpubidadjcol in the second
% column for its text to clear the IEEEpubid mark.

% use for special paper notices
%\IEEEspecialpapernotice{(Invited Paper)}

% make the title area
\maketitle

% As a general rule, do not put math, special symbols or citations
% in the abstract or keywords.
\begin{abstract}
In this letter, we propose a computationally efficient method for joint selection of cancellation carriers (CCs) and calculation of their values minimizing the out-of-band (OOB) power in non-contiguous (NC-) OFDM transmission. The proposed new CCs selection method achieves higher OOB power attenuation than algorithms known from literature as well as noticable reception performance improvement.   
\end{abstract}

% Note that keywords are not normally used for peerreview papers.

% For peer review papers, you can put extra information on the cover
% page as needed:
% \ifCLASSOPTIONpeerreview
% \begin{center} \bfseries EDICS Category: 3-BBND \end{center}
% \fi
%
% For peerreview papers, this IEEEtran command inserts a page break and
% creates the second title. It will be ignored for other modes.
\IEEEpeerreviewmaketitle

\section{Introduction}
\IEEEPARstart{P}{ractical} 
%spectrum shaping 
spectrum shaping algorithms are of great importance for the development of wireless communications systems as they allow for systems coexistence in the adjacent frequency bands and higher spectrum utilization. In case of non-contiguous orthogonal frequency division multiplexing (NC-OFDM) high out-of-band (OOB) radiation is observed at the output of the power amplifier (PA) caused by: high Peak-to-Average Power Ratio (PAPR) and related PA nonlinear distorions, as well as by the subcarrier spectrum sidelobes having the shape of the \emph{sinc}-like function. Although digital filtering can reduce the OOB radiation it involves
%it was shown in \cite{Yamaguchi04} that digital filtering involves 
high computational complexity  especially when high filter-selectivity is required \cite{Yamaguchi04}. 
Moreover, in the dynamically changing radio frequency opportunities, on-line filter design is usually unacceptable. 
Cancellation carriers method presented in \cite{Yamaguchi04,Brandes06} is a  flexible way of reducing OOB power by minimizing the spectrum sidelobes. It 
does not have limitations of filtering, and provides significant 
%out-of-band (
OOB power attenuation, especially when steep and narrow spectrum notches are required. Unfortunately, in some cases CCs can use significant fraction of the OFDM symbol power. In \cite{Brandes06,Huang09} optimization mechanisms for each OFDM symbol are suggested to overcome this problem. However, on-line optimization are computationally complex and impractical in hardware implementation. Extension of CCs utilizing subcarriers non-orthogonal to the OFDM subcarriers and frequency bins in the OOB region has been presented in \cite{EAIC-CP10}. The method results in high OOB radiation suppression at the cost of increased computational complexity and introduced self interference. An overview of the existing sidelobes suppression methods can be found in \cite{KryszkiewiczEURASIP12}.
% where the hardware-tested method allowing for simple CCs calculation for each OFDM symbol is also presented as well as 
There, the combination of the CCs and the time-domain windowing is discussed, and its performance results are presented showing higher OOB power suppression than the CCs method alone. 
%, and for the reception performance improvement thanks to the application of the time-independent precoding matrix. 
%For additional improvement in OOB performance, \cite{KryszkiewiczEURASIP12}. 
The drawback of the CC method is that it has to be adjusted for each particular subcarriers pattern matching fragmented frequency bands. 

In this letter, we show a computationally efficient method of the CCs calculation based on stochastic approach, which dynamically adjusts to the subcarriers-pattern. The mean CCs power is also constrained. The computational complexity of CCs calculation is decreased for each OFDM symbol in comparison with \cite{Brandes06} as it boils down to  matrix-vector multiplication. Additionally, determination of the CCs-calculation matrix is made less computationally expensive. The CCs-calculation matrix is constant for a given set of system parameters and can be used at the receiver for improving reception quality. 

Moreover, in this letter, the location of CCs is also revised. Typically equal number of CCs is placed on each side of data-occupied subcarriers blocks \cite{Brandes06,Yamaguchi04}. 
Here, we propose optimized CCs selection (OCCS), i.e. the heuristic approach to choose CCs locations iteratively. Our method outperforms significantly traditional approachs in terms of the OOB power reduction for a given number of CCs, what is obtained at the only cost of the iterative off-line low-computationally complex design of the CCs-calculation matrix.

In Section \ref{sec_2} we describe the system model. In Section \ref{sec_CCs} we state the optimization problem, and present its computationally-efficient solution and the OCCS heuristic. Simulation results are shown in Section \ref{sec_simulation} followed by the conclusions in Section \ref{sec_conculsion}.

\section{System model}
\label{sec_2}
We consider a wireless digital communication link based on NC-OFDM technique. In every OFDM symbol interval, $\alpha$ complex data symbols e.g. QAM, constituting a vector $\mathbf{d_{DC}}$ modulate the data carriers (DC) which are selected input bins of the $N$-th order Inverse Fast Fourier Transform (IFFT). Another set of $\beta$ subcarriers is used as CCs, and constite vector $\mathbf{d_{CC}}$. Merging of both vectors $\mathbf{d_{DC}}$ and $\mathbf{d_{CC}}$ results in vector $\mathbf{d}$ of length $\alpha+\beta\leq N$. Vectors of data and cancellation carriers indices are denoted as $\mathbf{I_{DC}}=\{I_{\mathrm{DC} j}\}$ and $\mathbf{I_{CC}}=\{I_{\mathrm{CC} l}\}$  respectively (where $j=1,...,\alpha$ and $l=1,...,\beta$). Merging of both vectors gives a vector $\mathbf{I_{C}}=[\mathbf{I_{CC}},  \mathbf{I_{DC}}]$ containing subcarriers indices from the set of $N$ IFFT inputs indexed as: 
$\{-N/2,...,N/2-1\}$. 
%We denote $I_{\mathrm{DC}}(j)$ as the $j$-th element in vector $\mathbf{I_{DC}}$. 
After IFFT, $N_{\mathrm{CP}}$ samples of cyclic prefix (CP) are inserted and symbol samples $y_{n}$ ($n=-N_{\mathrm{CP}},...,N-1$) are subject to the D/A conversion.
%, RF up-conversion and transmission.         

The OFDM symbol spectrum $S(v)$ at normalized frequency $v$ can be obtained by Fourier transformation of a time domain signal $y_{n}$ as in \cite{Yamaguchi04, KryszkiewiczEURASIP12}:
\begin{equation}
S(v)=\frac{1}{\sqrt{N}}\sum_{k\in \mathbf{I_{C}}}d_{k}S(v,k)\;\mathrm{,}
\label{eq_widmo3}
\end{equation}
where $S(v,k)$ is the $k$-th subcarrier spectrum:   
\begin{equation}
S(v,k)=\sum_{n=-N_{CP}}^{N-1}\exp\left(j2\pi\frac{n(k-v)}{N}\right).
\label{eq_widmo4}
\end{equation}

\section{New improved cancellation-carriers method}
\label{sec_CCs}
The idea of CCs is to modulate selected carriers with specific non-information symbols which minimize the OOB power measured at certain spectrum sampling points constituting vector $\mathbf{V}=\{V_{i}\}$ of $\gamma$ elements ($i=1,...,\gamma$). Given vectors 
$\mathbf{V}$ and $\mathbf{I_{DC}}$ we can define matrix $\mathbf{P_{DC}}=\{P_{\mathrm{DC} i,j}\}$ of size $\gamma\times\alpha$ with elements $P_{\mathrm{DC} i,j}=S\left(V_{i},I_{\mathrm{DC} j}\right)$. Similarly we can define matrix $\mathbf{P_{CC}}=\{P_{\mathrm{CC} i,l}\}$ of size $\gamma\times\beta$, where $P_{\mathrm{CC} i,l}=S\left(V_{i},I_{\mathrm{CC} l}\right)$. 
%for the CCs using vectors $\mathbf{V}$ and $\mathbf{I_{CC}}$. 
Then, the optimization problem to find vector $\mathbf{d_{CC}}$ resulting in the minimum of the OOB power:
\begin{align}
\min_{\mathbf{d_{CC}}}&\Vert \mathbf{P_{CC}d_{CC}}+\mathbf{P_{DC}d_{DC}}\Vert^{2}
\label{eq_optymalizacja1}
\\&s.t.~\Vert \mathbf{d_{CC}} \Vert^{2}\leq\beta \nonumber \;\mathrm{,}
\end{align}
where $\Vert \cdot \Vert$ is the Euclidean vector norm.
Here above, it is assumed that the power of CCs should be equal or lower than $\beta$ making mean CC power equal to or lower than the normalized data symbols power, which is assumed to be 1. 

\subsection{New efficient way of problem solution}
\label{sec_problem_solution}
Let us study the Lagrange function for problem (\ref{eq_optymalizacja1}):
\begin{equation}
f(\mathbf{d_{CC}},\theta)=\Vert \mathbf{P_{CC}d_{CC}}+\mathbf{P_{DC}d_{DC}}\Vert^{2}+\theta\left(\Vert \mathbf{d_{CC}}\Vert^{2}-\beta\right)	
\label{eq_lagrangian}
\end{equation}
where $\theta$ is the Lagrange multiplier active, i.e. taking the value higher than zero, when CCs power is to be limited and 0 otherwise, according to Karush-Kuhn-Tucker conditions. The solution of $\frac{\partial f(\mathbf{d_{CC}},\theta)}{\partial \mathbf{d_{CC}}}=0$ gives
\begin{equation}
\mathbf{d_{CC}}=-\left(\mathbf{P_{CC}}^{\mathcal{H}}\mathbf{P_{CC}}+\theta \mathbf{I} \right)^{-1}\mathbf{P_{CC}}^{\mathcal{H}}\mathbf{P_{DC}d_{DC}}=\mathbf{Wd_{DC}}
\label{eq_rozwiazanie}
\end{equation}
where $(~)^{\mathcal{H}}$ and $\mathbf{I}$ denote Hermitian transpose and identity matrix respectively, and $\mathbf{W}$ is our CCs-calculation matrix. Typically, as in \cite{Brandes06}, $\theta$ has to be found  to satisfy the constraint from (\ref{eq_optymalizacja1}) for each OFDM symbol. Here, we focus on satisfying the condition of the mean CCs power  assuming that random symbols in $\mathbf{d_{DC}}$ are independent with zero mean and unit variance. This mean CCs power equals:
 \begin{align}
\label{eq_expectation}
&\mathbb{E}[\|\mathbf{d_{CC}}\|^{2}]=
tr\left(\mathbb{E}\left[\mathbf{d_{DC}}^{\mathcal{H}}\mathbf{W}^{\mathcal{H}}\mathbf{W}\mathbf{d_{DC}}\right]\right)=
\\&\mathbb{E}\left[tr\left(\mathbf{d_{DC}}^{\mathcal{H}}\mathbf{W}^{\mathcal{H}}\mathbf{W}\mathbf{d_{DC}}\right)\right]=\nonumber
\\&tr\left(\mathbb{E}\left[\mathbf{d_{DC}}\mathbf{d_{DC}}^{\mathcal{H}}\right]\mathbf{W}^{\mathcal{H}}\mathbf{W}\right)
=tr\left(\mathbf{W}^{\mathcal{H}}\mathbf{W}\right)=\nonumber
\\&tr\left(\mathbf{P_{DC}}^{\mathcal{H}}\mathbf{P_{CC}}\left(\mathbf{P_{CC}}^{\mathcal{H}}\mathbf{P_{CC}}+\theta \mathbf{I} \right)^{-2}\mathbf{P_{CC}}^{\mathcal{H}}\mathbf{P_{DC}}\right)\nonumber \; \mathrm{,}
\end{align}
where $\mathbb{E}\left[~\right]$ and $tr(~)$ denote the expectation and the matrix trace, respectively. Expression (\ref{eq_expectation}) is obtained due to the linearity of trace and expectation operators and cyclic property of trace.   
As this equation is nonlinear, finding the value of $\theta$, for which  $\mathbb{E}\left[\mathbf{d_{CC}}^{\mathcal{H}}\mathbf{d_{CC}}\right]\leq \beta$  requires the use of the Newton method, where calculation of matrix inverse and a number of matrix-by-matrix multiplications is done in each iteration. 

Decreased computational complexity without results accuracy deterioration can be obtained by replacing $\mathbf{P_{CC}}$ by its singular value decomposition (SVD), i.e. $\mathbf{P_{CC}}=\mathbf{USV}^{\mathcal{H}}$, where $\mathbf{U}$ and $\mathbf{V}$ are unitary matrices, and $\mathbf{S}$ is $\gamma\times \beta$ diagonal matrix with $\delta$ singular values on its diagonal. Assuming full rank $\mathbf{P_{CC}}$ (what is usually the case), i.e. $\delta$ equal to the minimum of $\beta$ and $\gamma$ we obtain:
 \begin{align}
\label{eq_expectation2}
&\mathbb{E}[\|\mathbf{d_{CC}}\|^{2}]=
\\&tr\left(\mathbf{P_{DC}}\mathbf{P_{DC}}^{\mathcal{H}}\mathbf{US}\mathbf{V}^{\mathcal{H}}
\left(\mathbf{V}
\left(\mathbf{S}^{\mathcal{H}}\mathbf{S}+\theta \mathbf{I}\right)
\mathbf{V}^{\mathcal{H}}
\right)^{-2}\mathbf{V}\mathbf{S}^{\mathcal{H}}\mathbf{U}^{\mathcal{H}}\right)=
\nonumber
\\&tr\left(\mathbf{P_{DC}}\mathbf{P_{DC}}^{\mathcal{H}}
\mathbf{US}
\left(\mathbf{S}^{\mathcal{H}}\mathbf{S}+\theta \mathbf{I}\right)^{-2}
\mathbf{S}^{\mathcal{H}}\mathbf{U}^{\mathcal{H}}\right)\nonumber\;\mathrm{,}
\end{align}
where the properties of trace, unitary matrix and matrix inversion have been used to obtain this result.
With a modicum of algebra applied to (\ref{eq_expectation2}) we get 
 \begin{equation}    
tr\left(\mathbf{A}~
diag\left(\frac{|S_{1,1}|^{2}}{(\theta+|S_{1,1}|^{2})^2},...,\frac{|S_{\delta,\delta}|^{2}}{(\theta+|S_{\delta,\delta}|^{2})^2},0,...,0\right)
\right)\leq \beta
\label{eq_trace_result}
\end{equation}
where $diag()$ denotes diagonal matrix with diagonal entries given in brackets and $\mathbf{A}$ is the positive semidefinite Hermitian matrix defined as: $\mathbf{A}=\mathbf{U}^{\mathcal{H}}\mathbf{P_{DC}P_{DC}}^{\mathcal{H}}\mathbf{U}$. By analyzing the matrix and trace properties we finally obtain:
\begin{equation}    
\mathbb{E}\left[\|\mathbf{d_{CC}}\|^{2}\right]=
\sum_{i=1}^{\delta}\frac{A_{i,i}|S_{i,i}|^{2}}{(\theta+|S_{i,i}|^{2})^2}
\leq \beta\;\mathrm{.}
\label{eq_result_without_trace}
\end{equation}
For a given set of CCs and DCs only three matrix operations, i.e. one singular value decomposition and two matrix-by-matrix multiplications, and a few scalar-based Newton algorithm iterations have to be performed to find $\theta$. 

Additionally we can use matrices $\mathbf{A}$, $\mathbf{S}$ and $\mathbf{V}$ for calculation of the mean OOB radiation and the final CCs-calculation matrix. The OOB radiation power, being minimized in problem (\ref{eq_optymalizacja1}) can be reformulated by using (\ref{eq_rozwiazanie}) as follows:
  \begin{equation}
\min_{\mathbf{d_{CC}}}\Vert{ \mathbf{G}\mathbf{d_{DC}}}\Vert^{2}\;\mathrm{,}
\label{eq_minimalizacja_G}
\end{equation}
where 
 \begin{equation}    
\mathbf{G}=\mathbf{P_{CC}W}+\mathbf{P_{DC}}.
\label{eq_matrix_G}
\end{equation}
The mean OOB radiation power 
\begin{equation}    
P_{\mathrm{OOB}}=\frac{1}{\gamma}
\mathbb{E}\left[
\Vert{ \mathbf{G}\mathbf{d_{DC}}}\Vert^{2}
\right]
\label{eq_mean_OOB_trace}
\end{equation}
can be found by averaging the norm in (\ref{eq_minimalizacja_G}) over all possible $\mathbf{d_{DC}}$ vectors (by repeating operations presented in (\ref{eq_expectation})) for the number of spectrum sampling points $\gamma$.
By following similar routine as in derivation from (\ref{eq_expectation}) to (\ref{eq_result_without_trace}) we obtain:
 \begin{equation}    
P_{\mathrm{OOB}}=\frac{1}{\gamma}
\left(
\sum_{i=\delta+1}^{\gamma}A_{i,i}+\sum_{i=1}^{\delta}A_{i,i}
\left(
\frac{\theta}{\theta+|S_{i,i}|^{2}}
\right)^{2}
\right),
\label{eq_mean_OOB_without_trace}
\end{equation}
while for a basic system not using CCs, the OOB power equals:
%\begin{equation}    
$P_{\mathrm{OOB}}=\frac{1}{\gamma}
\|\mathbf{P_{DC}}\|^{2}$.
%\label{eq_mean_OOB_reference}
%\end{equation}
Finally, CCs calculation matrix can be obtained as:
 \begin{equation}    
\mathbf{W}=-\mathbf{V_{\delta}}diag\left(\frac{S_{1,1}^{\mathcal{H}}}{\theta+|S_{i,i}|^{2}}, ..., \frac{S_{\delta,\delta}^{\mathcal{H}}}{\theta+|S_{\delta,\delta}|^{2}}\right)\mathbf{U_{\delta}}^{\mathcal{H}}\mathbf{P_{DC}}\;\mathrm{,}
\label{eq_prec_matrix}
\end{equation}
where $\mathbf{V_{\delta}}$ and $\mathbf{U_{\delta}}$ are submatrices containing only first $\delta$ columns of matrices $\mathbf{V}$ and $\mathbf{U}$, respectively. The optimum $\mathbf{d_{CC}}$ vector can be now easily calculated using this matrix $\mathbf{W}$ as in formula (\ref{eq_rozwiazanie}). 

\subsection{Computational complexity}
\label{sec_complexity}
The computational complexity of the original CCs algorithm  is relatively high. According to \cite{Golub96} for each OFDM symbol about $\alpha \gamma+ \delta \beta+2\delta M$ operations are needed for the calculation of the CCs symbols, where $M$ is number of Newton method steps. In our case, (\ref{eq_rozwiazanie}) can be used 
directly or with precalculation of $\mathbf{P_{DC}d_{DC}}$ ($\mathbf{P_{DC}}$ decomposed from $\mathbf{W}$ for $\gamma \ll \beta$),    
%with or without precalculation of $\mathbf{P_{DC}d_{DC}}$, 
what  requires $\alpha \beta$ or $\alpha \gamma + \beta \gamma$ operations respectively. The computational complexity of our method in any case is significantly lower than the original one from \cite{Brandes06}. 
%In order to estimate computation speed-up on the stage of multiplier $\theta$ determination, let us compare complexity of calculation $\mathbb{E}\left[\|\mathbf{d_{CC}}\|^{2}\right]$ using (\ref{eq_expectation}) and (\ref{eq_result_without_trace}). In the first case no. computations order is $\beta^{3} +\alpha \beta +\alpha \beta \gamma$ (assuming matrix inversion using Gauss-Jordan elimination) while in the second it is only $\delta$. In both cases precomputation of matrices independent from $\theta$ is assumed.       

\subsection{Heuristic approach to CCs allocation scheme: OCCS}
\label{sec_heuristic}
\begin{figure}[!t]
\centering
\includegraphics[width=1\linewidth,draft=false]{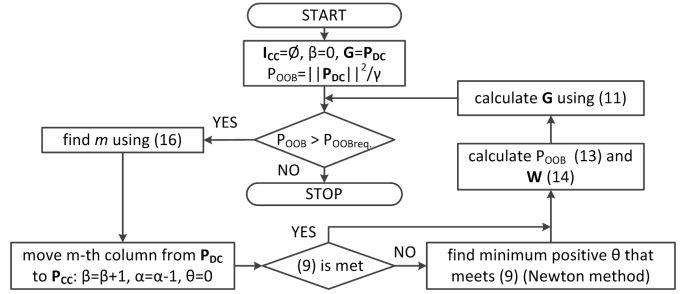}
\caption{Flow diagram of the OCCS algorithm.}
\label{fig_schemat_blokowy}
\end{figure}

Let us now review matrix $\mathbf{G}$ defined in (\ref{eq_matrix_G}). It projects data symbols $\mathbf{d_{DC}}$ onto spectrum samples in normalized frequencies $\mathbf{V}$ after CCs insertion. As data symbols are independent random variables with zero mean and unit mean power, coefficient $\left|G_{i,j}\right|^{2}$ is the mean OOB radiation power caused at frequency sampling point $V_{i}$ by data subcarrier indexed by $I_{\mathrm{DC} j}$. If we partition $\mathbf{G}$ into vertical vectors $\mathbf{g_{j}}$, i.e. 
$\mathbf{G}=[\mathbf{g_{1}}, \mathbf{g_{2}},...,\mathbf{g_{\alpha}}]$, the mean OOB power over all spectrum sampling points caused by  $I_{\mathrm{DC} j}$-th data subcarrier equals:
\begin{equation}   
\|\mathbf{g_{j}}\|^{2}=
\sum_{i=1}^{\gamma}\left|G_{i,j}\right|^{2}.
\label{eq_heurystyka3}
\end{equation} 
It is obvious that the data subcarrier causing the highest OOB power has the biggest influence on the OOB region. As such it has the highest potential to be used as CC to be added in the counterphase to strong OOB components caused by other subcarriers, i.e. to cancel them. If it is used as CC it should decrease some $|G_{i,j}|^{2}$ components. The OCCS criteria for finding index $m$ of the DC to be used as CC is the following: 
\begin{equation}   
m=\argmin_{j}\|\mathbf{g_{j}}\|^{2}.
\label{eq_heurystyka4}
\end{equation} 
The CCs selection has to be done iteratively as above. Single CC selection changes the matrix $\mathbf{W}$ changing also correlation properties between subcarriers. Typically, selection of a single CC causes that the other DCs in its neighborhood are not chosen to be the CC in the next step, because their influence on the OOB radiation is similar (highly correlated). The algorithm stops when the OOB radiation power achieves the required level $P_{\mathrm{OOB req}}$ (see the flow diagram of the OCCS in Fig.\ref{fig_schemat_blokowy}).

Finally note, that the OCCS can be combined with the time-domain windowing similarly as the standard CCs method \cite{KryszkiewiczEURASIP12}.

\section{Simulation results}
\label{sec_simulation}
For the proposed algorithm evaluation, we have considered an example NC-OFDM system with $N=256$ subcarriers spaced by 15kHz (as in the LTE system \cite{3gpp.36.101}) and a set of occupied subcarriers indexed as: $\mathbf{I_{C}}=\{-80,...,16\}\cup\{49,...,80\}$. 

\begin{figure}[!t]
\centering
\includegraphics[width=1\linewidth,draft=false]{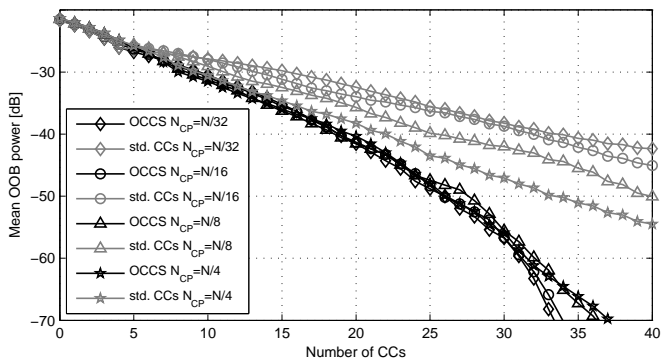}
\caption{The comparison of mean OOB power for standard (std.) CCs selection and OCCS scheme for different cyclic prefix durations at the PA input.}
\label{fig_OOB_vs_CCscheme}
\end{figure}

Notch spanning subcarriers from $17$ to $48$ (480kHz) can be occupied by a narrowband licensed system, e.g. a wireless microphone. 
The subcarrier indexed $0$ is unoccupied. The algorithm uses $\gamma=485$ spectrum sampling points $\mathbf{V}$ distributed equally in the normalized frequency region $\langle-125.75;-81\rangle \vee \langle 17;48\rangle \vee \langle 81;125.75 \rangle$. 
Matrix $\mathbf{W}$ is calculated for a number of CCs varying from 0 to 40, for various CP durations, for both standard CCs selection method and the OCCS heuristic. For the OCCS the heuristic algorithm in Fig. \ref{fig_schemat_blokowy} has been used. 
For the standard selection of CCs, 
the index $m$ indicates a subcarrier lying at the NC-OFDM band edges (not (\ref{eq_heurystyka4})).  
%At each stage 
%we used optimization method derived in Section \ref{sec_problem_solution} with application of Newton method for solving equation (\ref{eq_result_without_trace}).
% starting with $\theta=0$. 
The resulting mean OOB power vs. CP duration is presented in Fig. \ref{fig_OOB_vs_CCscheme}. Note, that while standard CCs selection needs different number of CCs to obtain given mean OOB power, the OCCS method is nearly independent from $N_{\mathrm{CP}}$. Moreover, for each number of CCs, the OCCS method is not worse (in terms of the mean OOB power) than the standard method while outperforming it significantly as the required OOB power level decreases. Additionally, the shorter the CP, the higher saving in the number of required CCs. For example, if we require the mean OOB power of at least $-40$dB, for $N_{\mathrm{CP}}=N/32$ we save  about 44\% of CCs (decrease from 34 to 19). These saved subcarriers can be used as DCs what increases the bit rate.  

\begin{figure}[!t]
\centering
\includegraphics[width=1\linewidth,draft=false]{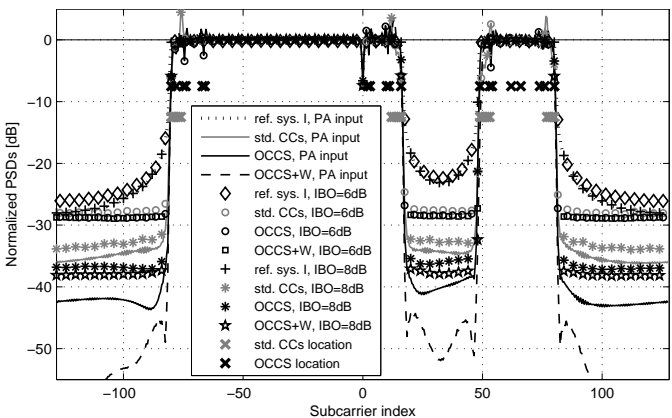}
\caption{PSDs of signals in ref. system I, the system with standard CCs, OCCS ($\beta=19$, $N_{\mathrm{CP}}=N/16$), and OCCS with windowing (W) ($\beta=15$, $N_{\mathrm{CS}}=10)$ at the PA input and output (with given IBO).} 
\label{fig_PSDs}
\end{figure}

For the remaining results presentation we have chosen the system with the fixed number of CCs $\beta=19$, $N_{\mathrm{CP}}=N/16$ and Gray mapped QPSK symbols. The comparison is done among the  NC-OFDM reference system I, i.e. without any spectrum shaping mechanism where all $\alpha+\beta$ subcarriers are DCs, and the systems applying standard CCs, our heuristic OCCS and OCCS combined with time-domain windowing (W) with $\beta=15$, and window cyclic suffix  $N_{\mathrm{CS}}=10$, i.e. the parameters resulting in the same bit rate as systems using CCs only. The Rapp-model of the PA has been used with nonlinearity hardness parameter $p=10$, and two values of the input back-offs (IBOs): 6 and 8 dB.
In Fig. \ref{fig_PSDs}, we can see the power spectral densities (PSDs) calculated over 10000 OFDM symbols. The locations of selected CCs are also marked. While the standard selection of CCs decreases OOB power by $12$~dB in comparison to reference system I, our OCCS method provides additional $6$~dB of the OOB power reduction. Moreover, our OCCS method lowers peaks rising in the in-band region, what helps the issue of satisfying the spectrum emission mask. Note that, as the IBO decreases, the intermodulations start to play dominant role in the OOB region and the effect of sidelobes minimization is less prominent. 

\begin{figure}[!t]
\centering
\includegraphics[width=1\linewidth,draft=false]{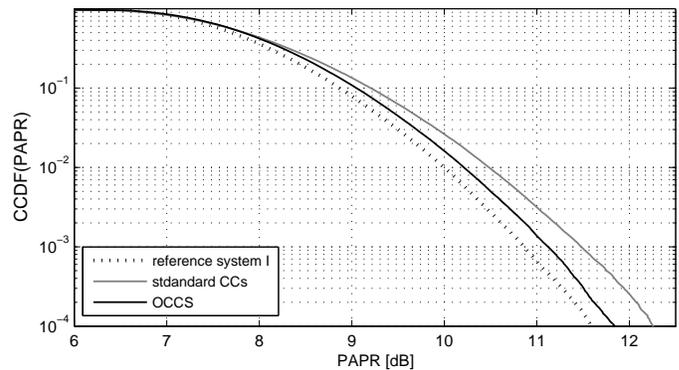}
\caption{CCDF of PAPR for NC-OFDM in the reference system I, and systems with standard and proposed CCs selection; $\beta=19$; $N_{\mathrm{CP}}=N/16$}
\label{fig_PAPR}
\end{figure}

Interestingly, our OCCS method results also in lower values of the complementary cumulative distribution function (CCDF) of PAPR than the standard CCs selection scheme, e.g. by  $0.4$~dB for $\mathrm{CCDF(PAPR)}=10^{-4}$, as shown in Fig.\ref{fig_PAPR}.   
%Peak-to-Average Power Ratio. These results are not presented here due to space limitations.         
% An example of a floating figure using the graphicx package.
% Note that \label must occur AFTER (or within) \caption.
% For figures, \caption should occur after the \includegraphics.
% Note that IEEEtran v1.7 and later has special internal code that
% is designed to preserve the operation of \label within \caption
% even when the captionsoff option is in effect. However, because
% of issues like this, it may be the safest practice to put all your
% \label just after \caption rather than within \caption{}.
%
% Reminder: the "draftcls" or "draftclsnofoot", not "draft", class
% option should be used if it is desired that the figures are to be
% displayed while in draft mode.
%

\begin{figure}[!t]
\centering
\includegraphics[width=1\linewidth,draft=false]{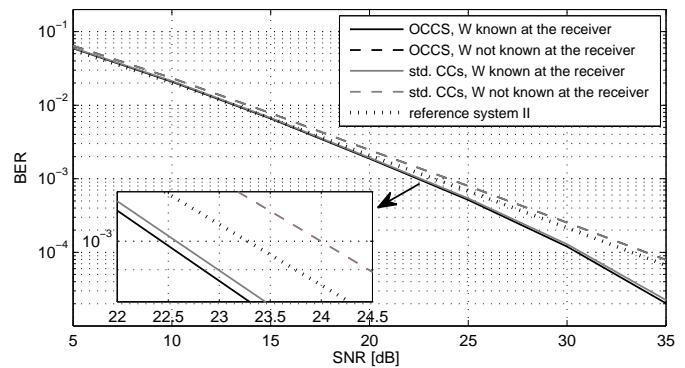}
\caption{BER vs. SNR for reference system II, the system with standard CCs alocation and OCCS; $\beta=19$, $N_{\mathrm{CP}}=N/16$}
\label{fig_BER}
\end{figure}

%The proposed OCCS scheme has also been investigated for its impact on the reception performance. 
In Fig. \ref{fig_BER}, the bit error rates (BERs) are shown for both CCs selection schemes considered with the option of matrix $\mathbf{W}$ available and nonavailable at the receiver. Both schemes are compared with the reference system II, in which subcarriers that could become standard CCs are instead modulated by zeros. This is for the sake of fair comparison of systems with equal bit rate. Matrix $\mathbf{W}$ can be used at the receiver to improve the performance making use of the redundancy symbols modulating CCs as shown in \cite{KryszkiewiczEURASIP12}. Here, the 9-paths Rayleigh channel model recommended for LTE \cite{3gpp.36.101} has been considered. There have been 50000 channel instances simulated with 1000 random OFDM symbols in each of them. The signal-to-noise ratio (SNR) loss with respect to the reference system II is the same for both CC selection methods when the $\mathbf{W}$ matrix is not known in the receiver. This is due to sacrificing some transmission power to the CCs. This SNR-loss equals $10\log_{10}\left(1+\beta/\alpha\right)$, i.e. about 0.7 dB in our scenario. The knowledge of the $\mathbf{W}$ matrix allows the receiver to decrease BER and obtain SNR gain even over reference system II (of 0.8 dB). The OCCS scheme outperforms the standard  approach to CCs selection, although improvement of BER performance is relatively small (0.15~dB for BER=$10^{-3}$). As the CCs improve mostly the reception quality of data subcarriers in their frequency neighborhood, the OCCS providing sparse CCs pattern causes more DCs to be positively influenced. 

\section{Conclusion}
\label{sec_conculsion}
The CCs calculation and OCCS algorithms presented in this letter allow for lower computational complexity, lower OOB power and lower PAPR in comparison with the standard CCs algorithm. Moreover, the OCCS approach increases the SNR gain at the receiver if the decoder makes use of the known CCs calculation (coding) matrix. These properties make the proposed OCCS algorithm superior over the existing ones.  

% if have a single appendix:
%\appendix[Proof of the Zonklar Equations]
% or
%\appendix  % for no appendix heading
% do not use \section anymore after \appendix, only \section*
% is possibly needed
% use appendices with more than one appendix
% then use \section to start each appendix
% you must declare a \section before using any
% \subsection or using \label (\appendices by itself
% starts a section numbered zero.)
%
% use section* for acknowledgement
%\section*{Acknowledgment}
% Can use something like this to put references on a page
% by themselves when using endfloat and the captionsoff option.
\ifCLASSOPTIONcaptionsoff
  \newpage
\fi

% trigger a \newpage just before the given reference
% number - used to balance the columns on the last page
% adjust value as needed - may need to be readjusted if
% the document is modified later
%\IEEEtriggeratref{8}
% The "triggered" command can be changed if desired:
%\IEEEtriggercmd{\enlargethispage{-5in}}

% references section

% can use a bibliography generated by BibTeX as a .bbl file
% BibTeX documentation can be easily obtained at:
% http://www.ctan.org/tex-archive/biblio/bibtex/contrib/doc/
% The IEEEtran BibTeX style support page is at:
% http://www.michaelshell.org/tex/ieeetran/bibtex/
\bibliographystyle{IEEEtran}
% argument is your BibTeX string definitions and bibliography database(s)
\bibliography{pawla_bib}
%
% <OR> manually copy in the resultant .bbl file
% set second argument of \begin to the number of references
% (used to reserve space for the reference number labels box)
%\begin{thebibliography}{1}

%\end{thebibliography}

% biography section
% 
% If you have an EPS/PDF photo (graphicx package needed) extra braces are
% needed around the contents of the optional argument to biography to prevent
% the LaTeX parser from getting confused when it sees the complicated
% \includegraphics command within an optional argument. (You could create
% your own custom macro containing the \includegraphics command to make things
% simpler here.)
%\begin{IEEEbiography}[{\includegraphics[width=1in,height=1.25in,clip,keepaspectratio]{mshell}}]{Michael Shell}
% or if you just want to reserve a space for a photo:

%\begin{IEEEbiography}{Michael Shell}
%Biography text here.
%\end{IEEEbiography}

% if you will not have a photo at all:
%\begin{IEEEbiographynophoto}{John Doe}
%Biography text here.
%\end{IEEEbiographynophoto}

% insert where needed to balance the two columns on the last page with
% biographies
%\newpage

%\begin{IEEEbiographynophoto}{Jane Doe}
%Biography text here.
%\end{IEEEbiographynophoto}

% You can push biographies down or up by placing
% a \vfill before or after them. The appropriate
% use of \vfill depends on what kind of text is
% on the last page and whether or not the columns
% are being equalized.

%\vfill

% Can be used to pull up biographies so that the bottom of the last one
% is flush with the other column.
%\enlargethispage{-5in}

% that's all folks
\end{document}